\documentclass[twocolumn,a4paper,nofootinbib,
showpacs,aps,floatfix,prl,superscriptaddress]{revtex4}
\usepackage{graphicx}
\usepackage{bm}
\usepackage{amsmath}
\newcommand{\ignore}[1]{}

\def\beq{\begin{equation}}
\def\eeq{\end{equation}}
\def\la{\langle}
\def\ra{\rangle}
\newcommand {\bp}{{\bf{p}}}
\newcommand {\bx}{{\bf{x}}}
\newcommand {\si} {/\mbox{s}}
\newcommand {\mum}{\, \mu \mbox{m}}
\newcommand {\cms}{\, \mbox{cm/s}}
\newcommand {\fexp} [1] {\exp \left( #1 \right)}
\newcommand {\fabs}[1] {\left| #1 \right|}
\newcommand {\fabsq}[1] {\left| #1 \right|^2}
\newcommand {\ms} {\, \mbox{ms}}
\begin{document}
\title{Atom cooling with an atom-optical diode on a ring} 
\author{A. Ruschhaupt}
\email[Email address: ]{a.ruschhaupt@tu-bs.de}
\affiliation{Institut f\"ur Mathematische Physik, TU Braunschweig, Mendelssohnstrasse 3, 38106 Braunschweig, Germany}
\author{J. G. Muga}
\email{jg.muga@ehu.es}
\affiliation{
$^b$Departamento de Qu\'imica-F\'isica, Universidad del Pais Vasco
  48080 Bilbao, Spain}

\begin{abstract}
We propose a method to cool atoms on a ring
by combining an atom diode --a laser valve for one-way atomic motion
which induces robust internal state excitation-- and a trap.
We demonstrate numerically that 
the atom is efficiently slowed down at each diode crossing, and it is 
finally trapped when its velocity is below the trap threshold. 

\end{abstract}

\maketitle

There is currently much interest in controlling the motion of cold atoms
for further (deeper) cooling, quantum information processing, atom laser generation,
metrology, interferometry, 
and the investigation of fundamental physical phenomena. 
Cold atoms are relatively easy to produce and offer with respect to other
particles many possibilities for coherent external manipulation with lasers,
magnetic fields, or mechanical interactions. They may be trapped in artificial
lattices or even  individually,
can be guided in effectively one-dimensional wires, or adopt interesting
collective behavior in Bose-Einstein condensates;
also, their mutual interactions can be changed, or suppressed. 
All this flexibility facilitates the translation of some of the concepts and
applications of electronic circuits into the atom-optical realm to implement
atom chips, atom circuits, or quite 
generally ``atomtronics'' \cite{atomtronics}.     
In this context, efficient elementary circuit elements playing the 
role of diodes or transistors need to be developed. In particular, 
we have proposed and studied a laser device acting as a one-way barrier 
for atomic motion \cite{RM04,RM06,RMR06q,RMR06d,RM07},
and similar ideas, near experimental verification, 
have been considered by
Raizen and coworkers for atom cooling \cite{raizen05,dudarev05}.
(These one-way models rely on atom-laser
interactions in the independent atom regime,  
but there are also complementary proposals making use of interatomic interaction for ``diodic'' 
one-way transport \cite{diodeH}.)
In the following we propose to combine, within a ring, 
the diode and a trap, to achieve cooling and trapping with phase-space
compression.
Ring-shaped traps for cold atoms have been proposed or implemented for 
matter-wave interferometry and highly precise sensors \cite{Gupta},
for studying the stability of 
persistent currents \cite{JPY,sombrero}, sound waves, 
solitons and vortices in Bose-Einstein
condensates \cite{vortex}, collisions \cite{Marcassa},
for coherent acceleration \cite{acceleration,BK}, 
production of highly directional output beams \cite{SBC}, 
or quantum computation \cite{qc}. (For further applications see \cite{Nugent}.)  
The ring traps are implemented by magnetic waveguides \cite{SBC,Gupta,Spreeuw},
purely optical dipole forces \cite{optical ring}, magnetoelectrostatic 
potentials \cite{me}, overlapping of magnetic and optical dipole traps 
\cite{opma}, or misalignment of counterpropagating 
laser beam pairs in a magneto-optical trap \cite{Marcassa}.   

% ---------------- FIG. 1 BEGINS ----------------
\begin{figure}[h]
\begin{center}
\includegraphics[angle=0,width=0.8\linewidth]{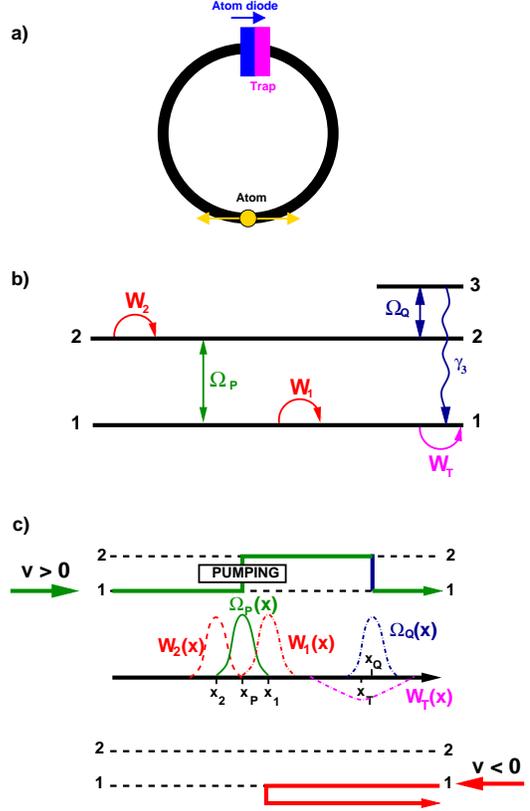}
\end{center}
\caption{\label{fig1}(a) Setting of an atom diode in a ring,
(b) schematic action of the different lasers 
on the atom levels for the two-level atom diode plus quenching, and 
(c) schematic spatial location of the different laser potentials
and their effect on the moving atom.
They are all taken as Gaussian functions:  
$\Omega_P(x) = \hat{\Omega}_P \Pi(x,x_P,\sigma)$, 
$W_\alpha(x) = \hat{W}_\alpha\; \Pi (x,x_{\alpha},\sigma_\alpha)$, 
where $\alpha=1,2,T,Q$;  $\sigma_{1,2}=\sigma$;  
and  
$\Pi (x, x_0, \tilde\sigma)= \fexp{-\frac{(x-x_0)^2}{2 {\tilde\sigma}^2}}$.
}
\end{figure}
% ---------------- END FIG. 1 ----------------
%

In this paper, we assume for simplicity tight lateral confinement
so that motion along the
ring is effectively one dimensional or, more exactly, circular
with a length $l$.
An atom diode followed by a trap
are put on the ring (see Fig. \ref{fig1}a). 
The initial state of ground-state atoms will have some  velocity 
and position width, but 
the anti-clockwise moving atoms will
be reflected by the diode -which can be
crossed only in the direction of the arrow- so all atoms 
will
eventually approach the diode clockwise.
After each crossing of the diode, the atom, which is now excited,   
is forced to emit a spontaneous photon (quenched),
and ends up at the bottom of the trap; the atom has to loose kinetic energy to escape
from the trap, so it is slowed down at every crossing
until being finally trapped when its velocity is below the
trap threshold.
The process is reminiscent of Sysiphus cooling \cite{sc}, a difference  
being that both the transfer from ground to 
excited state in the diode and the quenched decay are
highly controlled, robust and efficient processes.   

\begin{figure}[t]
\begin{center}
\includegraphics[angle=0,width=\linewidth]{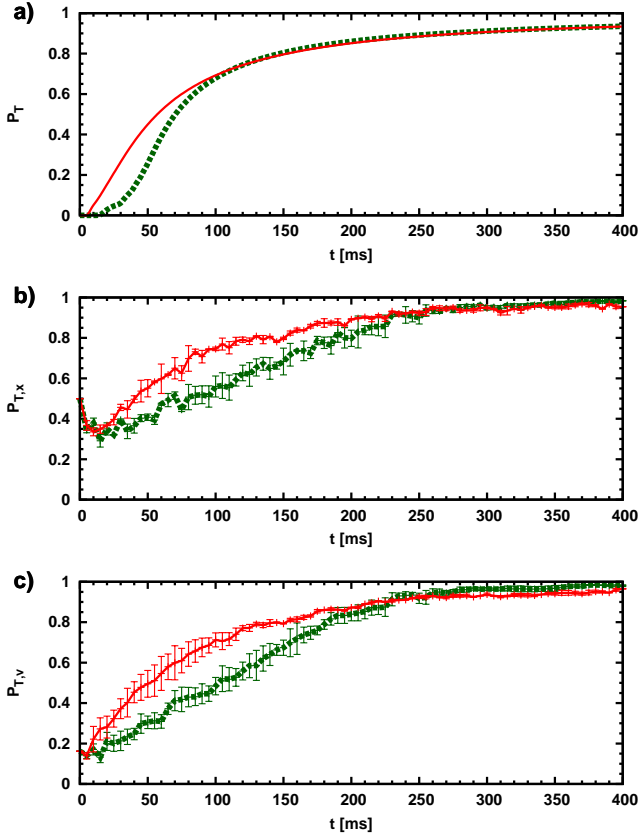}
\end{center}
\caption{\label{fig2} 
(a) Trapping probability in the classical model; 
(b) trapping probability $P_{T,x}$ in the quantum model
($x_{min}=10\mum$ and $x_{max}=200\mum$); and 
(c) trapping probability $P_{T,v}$ in the quantum model. 
Thick, green, dotted line: $v_{rec}=0$
(in the quantum model:
averaged over $N=200$ trajectories,
$\hat{\Omega}_P = 4\times 10^4\si$,
$\hat{W}_1=\hat{W}_2=4\times 10^6\si$);
solid, red line: $v_{rec}=3.5 \cms$
(in the quantum model: averaged over $N=190$ trajectories,
$\hat{\Omega}_P = 1\times 10^5\si$,
$\hat{W}_1=\hat{W}_2=1\times 10^7\si$).
Common parameters:
$l = 400 \mum$ ($-200 \mum \le x < 200\mum$),
$m$ mass of Neon,
initial wave packet
$
\Psi_0 (x) = \frac{1}{\sqrt{2\pi}} \int dk\, \Phi_0 (k) e^{i k x}
$
with
$\Phi_0 (k) = \frac{1}{\sqrt[4]{2\pi}\sqrt{\Delta k}}
(1,0,0)^T
\exp\Big[-\frac{(k-k_0)^2}{4 \Delta k^2} - i (k-k_0) (x_0-\frac{\hbar}{m} t_0 k_0)
- i \frac{\hbar}{m} t_0 \frac{k^2}{2}\Big] 
$, $k_0 = \frac{m}{\hbar} v_0$, $\Delta k = \frac{m}{\hbar} \Delta v$,
$x_0=-200\mum$, $v_0=5 \cms$, $\Delta v= 4\cms$, $t_0=1\ms$;
other parameters in the classical model:
$v_{T} = 1.8 \cms$, $x_D = 80 \mum$;
other parameters in the quantum model:
$\hat{W}_{T} = -10^5\si$,
$\hat{W}_{Q} = 10^5\si$,
$x_{W2} = -90 \mum$, $x_P = -40\mum$, $x_{W1}=10\mum$,
$x_{T}=80\mum$, $x_{Q} = 100 \mum$, 
$\sigma = 15 \mum$,
$\sigma_{T}=30\mum$, $\sigma_{Q}=10/\sqrt{2}\mum$.
}
\end{figure}

Before looking at the quantum-mechanical description,
we will examine a simple classical model 
to estimate the time
scales and the cooling efficiency for different recoil velocities.
In this classical toy model, the diode and the trap are reduced to a point
at position $x_D$, and the initial particle   
positions and momenta are distributed according to Gaussian probability
distributions corresponding to the initial quantum
distributions $\fabsq{\Psi_0(x)}$ resp. $\fabsq{\Phi_0 (k)}$
(see below).
In each classical trajectory a random recoil kick is imparted
at the diode clockwise passage, as in the quantum jump calculation done below.
The trajectory is ``trapped'' (and eliminated from the ensemble) 
when the velocity becomes smaller than the threshold imposed by the trap
depth; otherwise a fixed amount of kinetic energy corresponding to the
well depth $\frac{1}{2}m v_T^2$ is subtracted and the motion continues.   
The results for the trapping probability are
shown in Fig. \ref{fig2}a.
Random recoil affects the result in two ways: 
higher recoil velocities accelerate a rapid initial trapping, 
but slightly increase the time necessary for cooling and trapping the 
complete ensemble. 
The number of diode crossings before the atom is trapped for 
an initial velocity $v>0$ and no recoil is given by
the smallest integer $n$ fulfilling $v - n v_T < v_T$. The
time until this particle has been trapped is given by the time
to reach the diode the first crossing, $t_0 = -(x_0 - x_D)/v$
plus the total time for the $n$ rounds,
$t_n = l \sum_{j=1}^n\left(v - j v_T\right)^{-1}$. For
the parameters of Fig. \ref{fig2}a and $v=v_0$
we get $n=2$ and $t_0+t_n \approx 41\ms$.

Now we switch to the quantum mechanical description.
The scheme of the diode used here can be seen in Fig. \ref{fig1}b and c
and it has been
explained before \cite{RM04,RM06,RMR06q,RM07}.
In brief, there are three, partially overlapping laser regions:
two of them are state-selective mirrors blocking the 
excited (level 2) and ground (level 1) state on the left and right,
respectively 
of a central pumping region on resonance with the atomic transition.
If the atom is traveling from the right and the velocity is not too
high,  it is reflected by the state-selective mirror
potential $W_1 \hbar/2$. 
If the atom is traveling from the left in the ground state
then it will be pumped to the second level
adiabatically (so that the process is robust and independent of velocity in
a broad velocity interval) 
and then pass the potential $W_1 \hbar/2$.
Note that this setting, see Fig. \ref{fig1}b, can be realized
by a detuned STIRAP transfer \cite{RM06} with just two overlapping lasers. 
To avoid backwards motion after the atom has crossed the diode
we assume a third level which decays to the ground state with a
decay rate $\gamma_3$ and we add a quenching laser
coupling levels 2 and 3 with a Rabi frequency $\Omega_Q$,
see Fig. \ref{fig1}b.
A novelty with respect to previous diode models is the addition of 
a ground state well
overlapping partially with   
the quenching laser region. The effect of this well is twofold: it subtracts
kinetic energy from 
the ground state atoms trying to escape from it, and it also traps eventually 
the cooled atoms.  
%
%
% ---------------- FIG. 2 BEGINS ----------------
% ---------------- END FIG. 2 ----------------
%
%
The corresponding Hamiltonian using $|1\rangle=(1,0,0)^T$, 
$|2\rangle=(0,1,0)^T$, and $|3\rangle=(0,0,1)^T$, where $T$ means
``transpose'',       
may now be written as  
\begin{eqnarray*}
H = \frac{\bp_x^2}{2m} + 
\frac{\hbar}{2} \left(\begin{array}{ccc}
W_1(x) + W_{T} (x) & \Omega_P (x) & 0\\
\Omega_P (x) & W_2(x) & \Omega_Q (x)\\
0 & \Omega_Q (x) & 0
\end{array}\right),
\end{eqnarray*}
where 
$\bp_x$ is the momentum operator,   
and $\Omega_P(x)$ is the Rabi
frequency for the resonant transition.  
All potentials are chosen as Gaussian functions according to the caption of Fig. 1. 
The ``velocity depth'' of the trap used is
$
v_{T} := \sqrt{\frac{\hbar}{m} \fabs{\hat{W}_{T}}}
\approx 1.8 \cms, % 1.774 \cms
$
and it corresponds to the trap depth used in the classical
simulation.
\begin{widetext}
We examine the time evolution by means of 
a one-dimensional master equation, 
\begin{eqnarray}
\frac{\partial}{\partial t} \rho
&=& - \frac{i}{\hbar} [H, \rho]_-
- \frac{\gamma_3}{2} \{|3\ra \la 3|,\rho\}_+ %\nonumber\\
+ \gamma_3 \int_{-1}^{1} \!\!du\; \frac{3}{8} (1+u^2) \;
\fexp{i\frac{mv_{rec}}{\hbar} u \bx}
\,|1\ra \, \la 3|\rho|3\ra \, \la 1|\,
\fexp{-i\frac{mv_{rec}}{\hbar} u \bx},
\label{master_eq}
\end{eqnarray}
where $v_{rec}$ is the recoil velocity and $\bx$ is the position
operator.
The initial condition is taken as a pure state
$\rho(0)=|\Psi_0\ra\la\Psi_0|$, namely a
Gaussian wave packet (see the caption of Fig. 2).  
\end{widetext}
The master equation (\ref{master_eq}) is solved by using the quantum jump
approach \cite{qjump}. A basic step is to solve a time-dependent Schr\"odinger
equation with
an effective Hamiltonian $H_{eff} = H - i \frac{\hbar}{2} \gamma_3 |3\ra \la 3|$.
For large $\gamma_3$  
(see \cite{ruschhaupt_2004_eeqt}), 
\begin{eqnarray}
\la 3|\Psi(t)\ra \approx -i \frac{\Omega_Q (x)}{\gamma_3} \la 2|\Psi(t)\ra.
\label{approx}
\end{eqnarray}
Therefore, the three-level Schr\"odinger equation can be approximated 
by a two-level one with the effective Hamiltonian
\begin{eqnarray*}
H_{approx} = \frac{\bp_x^2}{2m} + 
\frac{\hbar}{2} \left(\begin{array}{cc}
W_1(x) + W_{T} (x) & \Omega_P (x)\\
\Omega_P (x) & W_2(x)-i W_{Q} (x)
\end{array}\right),
\end{eqnarray*}
where $W_{Q} = \frac{\Omega_Q (x)^2}{\gamma_3}$.
The second element
of the approach is the resetting operation at each jump, 
$
\fexp{i\frac{mv_{rec}}{\hbar} u \bx} \la 3|\Psi(t) \ra \longrightarrow \la 1|
\Psi(t)\ra,
$
where $u \in [-1,1]$ is chosen with the probability density $\frac{3}{8}
(1+u^2)$, all other amplitudes are set to zero, and the wave function
is normalized. Because of Eq. (\ref{approx}), this can also be done
in the two-level approximation, 
$
-i \sqrt{W_{Q} (\bx)} \fexp{i\frac{mv_{rec}}{\hbar} u \bx} \la
2|\Psi(t) \ra \longrightarrow \la 1|
\Psi(t)\ra,
$
then the second level is set to zero and the wave function is normalized.
%
% ---------------- FIG. 3 BEGINS ----------------
\begin{figure}[t]
\begin{center}
\includegraphics[angle=0,width=\linewidth]{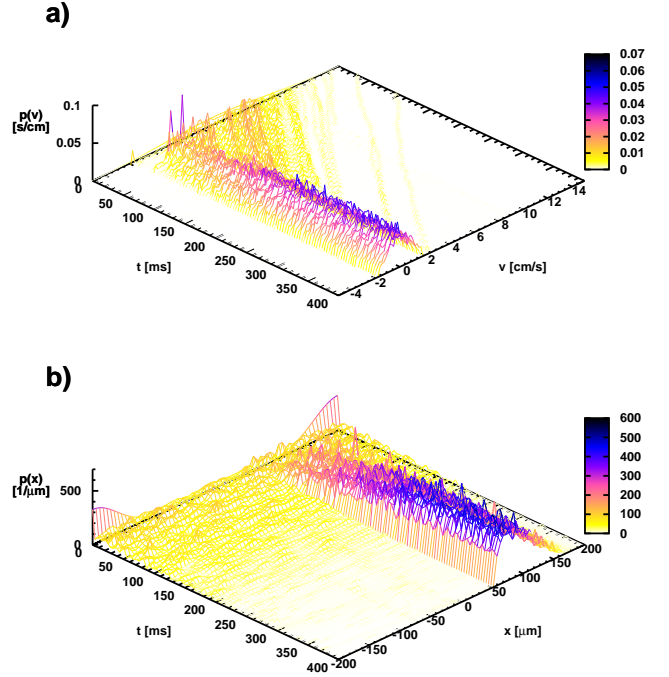}
\end{center}
\caption{\label{fig3} Evolution of the probability densities versus time
(a) in velocity space, (b) in coordinate
space; $v_{rec}=0$, the other parameters can
be found in the caption of Fig. \ref{fig2}.}
\end{figure}
% ---------------- END FIG. 3 ----------------

We start by looking at the case with negligible 
recoil velocity, i.e. with $v_{rec}=0$.
We calculate the  
trapping probability in coordinate space, 
$
P_{T,x} = \int_{x_{min}}^{x_{max}} dx \, \la x|\rho_{11}|x\ra,
$
and
in velocity space, 
$
P_{T,v} = \int_{-v_{T}}^{v_{T}} dv\, \la v|\rho_{11}|v\ra.
$
The results are shown in Fig. \ref{fig2}b/c (thick green
dotted line) averaging over $N=200$ trajectories; a numerical error
defined by the difference of the result between averaging
over $N$ and $N/2$
trajectories is also plotted in Fig. \ref{fig2}b/c.
The parameters used for the atom diode with $v_{rec}=0$ result
in a range for perfect ``diodic'' behavior
$-11 \cms < v < 11 \cms$
(defined as in \cite{RM06}).
%
%
% ---------------- FIG. 4 BEGINS ----------------
\begin{figure}[t]
\begin{center}
\includegraphics[angle=0,width=7.5cm]{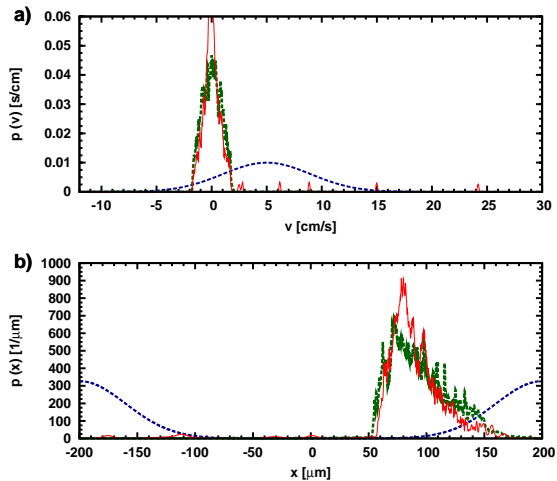}
\end{center}
\caption{\label{fig4} Initial probability density (dotted blue line);
final probability density at $t=400 \ms$ for $v_{rec} = 0$
(thick, green, dotted line),
$v_{rec} = 3.5 \cms$  (solid, red line);
other parameters see Fig. \ref{fig2};
(a) in velocity space; (b) in coordinate space.}
\end{figure}
% ---------------- END FIG. 4 ----------------
%
Fig. \ref{fig3} shows the evolution of the probability density
in velocity space, $p(v) = \la v | (\rho_{11} + \rho_{22}) | v\ra$,
and in coordinate space,
$p(x) = \la x | (\rho_{11} + \rho_{22}) | x\ra$.
The final probability densities
can be seen in more detail in Fig. \ref{fig4} (thick, dashed, green
line). In this figure we may verify the occurrence of 
cooling and phase space compression, namely, 
a narrower distribution both in coordinate and velocity space.  
The final trapping probabilities are
$P_{T,x} = 0.9838 \pm 0.0009$ and
$P_{T,v} = 0.9809 \pm 0.0040$
(with the errors calculated as before). 

Let us now examine the case with recoil velocity
$v_{vec}=3.5 \cms > v_{T}$.
Because the average value of the recoil velocities  
is still zero,
the cooling method will still work except for a small 
fraction of atoms which may acquire by successive random recoils 
velocities above the break-up threshold of the diode.
(The parameters used for the atom diode with $v_{rec}=3.5\cms$ result
in a working range $-17.5 \cms < v < 22 \cms$.)
The trapping probability versus time shown
in Fig. \ref{fig2}b (solid, red line) shows anyway  
a high final trapping probability. 
The final probability densities are shown in Fig. \ref{fig4}
(solid, red line). The main peak is comparable with the main
peak without recoil, i.e. the velocity width of the main peak
is smaller than the recoil velocity.
We have finally
$P_{T,x} = 0.9544 \pm 0.0003$
and $P_{T,v} = 0.9654 \pm 0.0017$.

In summary, we have proposed and numerically demonstrated 
a method to cool atoms on a ring, even below recoil velocity, 
after repeated passages across an atom diode combined with a ground
state trap.  

\begin{acknowledgments}
We acknowledge ``Acciones Integradas'' of the German
Academic Exchange Service (DAAD) and of Ministerio de
Educaci\'on y Ciencia (MEC). 
This work has also been supported by MEC
(FIS2006-10268-C03-01) and UPV-EHU (00039.310-15968/2004).
AR acknowledges support by the Joint Optical Metrology Center (JOMC),
Braunschweig. 
\end{acknowledgments}

\end{document}